# THE EMERGENCE OF COGNITIVE DIVERSITY IN IDEA MARKETS

Paul Dwyer[1]

## ABSTRACT

Idea markets are contexts where ideas compete for attention and people to embrace them. They are near ubiquitous in the form of religions, political parties, securities markets and the blogosphere to name a few. Most idea markets are also communities where members alternate between being idea consumers and producers. They are thus centers for collaborative knowledge creation. Recent research extols the value of diverse perspectives among the members of such groups, however these groups tend to form based on commonality of idea preference, a basis that is thought to limit diversity. This study investigates whether ideal levels of cognitive diversity can emerge in such groups. It finds that general popularity among idea markets draws together people with diverse perspectives, causing ideal levels of cognitive diversity to emerge.

## INTRODUCTION

A Google search of the term *idea market* yields a variety of different concepts associated with the term, ranging from small face-to-face brainstorming gatherings to what is more commonly called *information* or *predictive markets*. Predictive markets are true speculative markets that trade *idea futures*, an asset whose monetary value is usually tied to the likelihood of an event, like the election of a presidential candidate or the success of a new product.

There is, however, a need for a term like "idea market" to describe a general context where ideas compete for people to embrace them, not in the sense of predicting whether they will eventually be shown to be true, but in the sense of being values, philosophies, opinions, beliefs and sentiments that are currently held as having value or meaning. Examples include the competition between religions, political parties, advertisements, books and weblogs for attention in open market systems. An idea market therefore exists to bring holders of idea preferences together with idea producers.

Many idea markets, such as religions, political parties, and weblogs, are also idea communities where members alternate between being idea consumers and idea producers. This calls attention to the dual nature of cognitive diversity: diverse preferences and diverse thinking styles. Recent research extols the virtues of a multiplicity of thinking styles in groups of all kinds as diverse thinking leads to better collaborative outcomes. However, it is uncertain whether the dynamics of group formation in idea markets, the uniting of people with similar idea preferences, will result in a group with a diversity of thinking styles that is able to collaborate and go beyond merely embracing an idea to extending it.

[1] Assistant Professor of Marketing, Atkinson Graduate School of Management, Willamette University, 900 State Street, Salem, OR 97301. This work was partially supported by the Santa Fe Institute through NSF Grant No. 0200500 entitled "A Broad Research Program in the Sciences of Complexity."

This study focuses on investigating the emergence of cognitive diversity with the following research questions: Is there a limit to the utility of cognitive diversity? Will an idea market facilitate the emergence of ideal levels of cognitive diversity? What causes ideal levels of cognitive diversity to emerge? How can cognitive diversity be measured? To address these questions this study is organized as follows: first, the primary literature that relates to collective cognition is surveyed. Then, several variations of an agent-based model are proposed to simulate how agents may interact in an idea market. The results of these models are then compared to each other, and to observations of a sample of weblogs, drawn from one of the most active and diverse idea markets: the blogosphere.

## OVERVIEW OF RELEVANT LITERATURE

### Idea Markets

A Google search of the term *idea market* yields a variety of different concepts associated with the term, ranging from small face-to-face brainstorming gatherings to Kambil's (2003) description of what is more commonly called an *information* or *predictive market*. Predictive markets are true speculative markets that trade *idea futures*, an asset whose monetary value is usually tied to the likelihood of an event, like the election of a presidential candidate or the success of a new product.

There is, however, a need for a term like "idea market" to describe a general context where ideas compete for people to embrace them, not in the sense of predicting whether they will eventually be shown to be true, but in the sense of being values, philosophies, opinions, beliefs and sentiments that are currently held as having value or meaning. One example is Warner's (1993) contention that a new paradigm recently emerged among U.S. religious institutions that force them to compete for members in an open market system rather than as near monopolies. He argues that individualistic tendencies among Americans cause them to seek a church that is a best fit to their lifestyles and values rather than one that honors a legacy from their familial or cultural past.

However, it is not just religious ideology that competes in an open market system. Since the Internet went mainstream people have used the enhanced connectivity to join and form online communities such as forums and weblogs, bringing together likeminded people that would never have met face-to-face. While search engines have been developed to help people find communities that closely match their interests, people also engage in a wandering process, sometimes aided by word-of-mouth referral from similar others, to find online communities wherein they feel most intellectually at home. Some seekers, passionate about their interests, have started their own virtual communities when their search for an ideal existing one has been unsuccessful.

It has been noticed by many internet researchers (e.g., Shirky 2003) that the populations associated with virtual communities follow a *power law* distribution such as that shown in Figure 1. Webb (2002) demonstrated that "power laws are characteristic of randomly distributed values that come from a scarce resource" and Shirky (2003) opined that in the blogosphere the scarce resource was people's time. People can only invest a limited amount of time searching for the ideal idea source. As a result they use means of referral, like search engines, that typically direct them to the most popular content that matches their search criteria. The effect of search engines is similar to Simon's (1955) finding that *preferential attraction* creates power law behavior: people are attracted to that which is already most popular. Limited time and populist influences thus create and sustain a power law distribution in idea markets.

**FIGURE 1**

**Power Law Profile of Idea Markets**

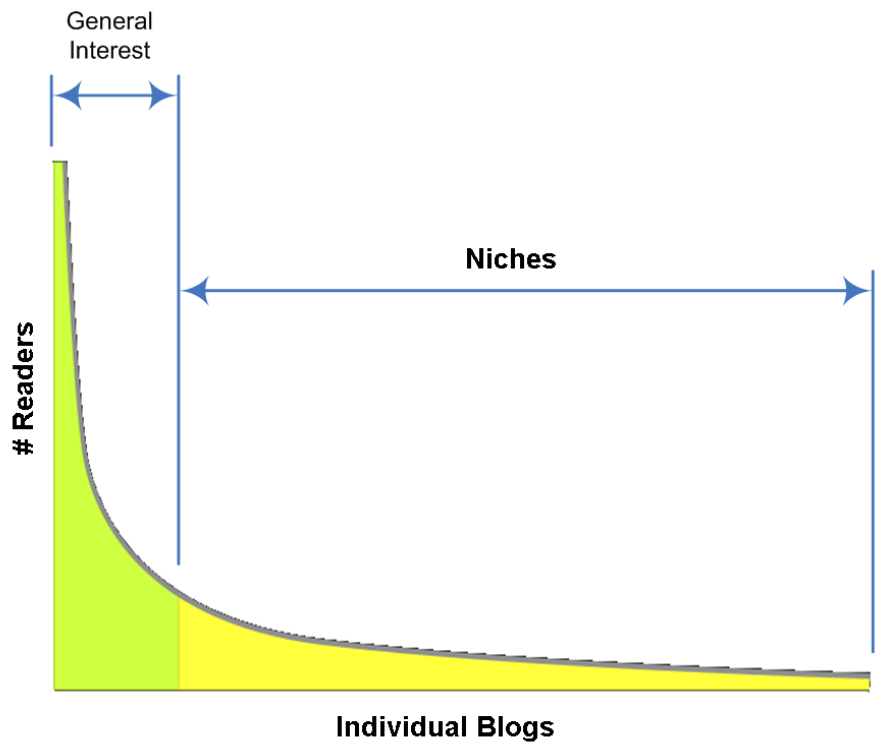

**Cognitive Diversity**

In the previous section we described how a scarcity of time and populist influences draws a disproportionate number of idea-seekers to a select few idea sources. It is understandable to think these prominent idea sources would be very influential, however Blaser (2007), a prominent blogger and entrepreneur, opines that its not passive readership that confers true influence but active discussion, and that most of the active discussion occurs in the *long tail* of the power law, among the people who took the time to find idea sources that truly engage their minds. He calls this *The People Law* and used Figure 2 to illustrate his point. Blaser believes that engaged minds will go beyond merely embracing ideas and extend them, achieving true collaborative learning and new knowledge creation as a result.

**FIGURE 2**

**Britt Blaser's People Law**

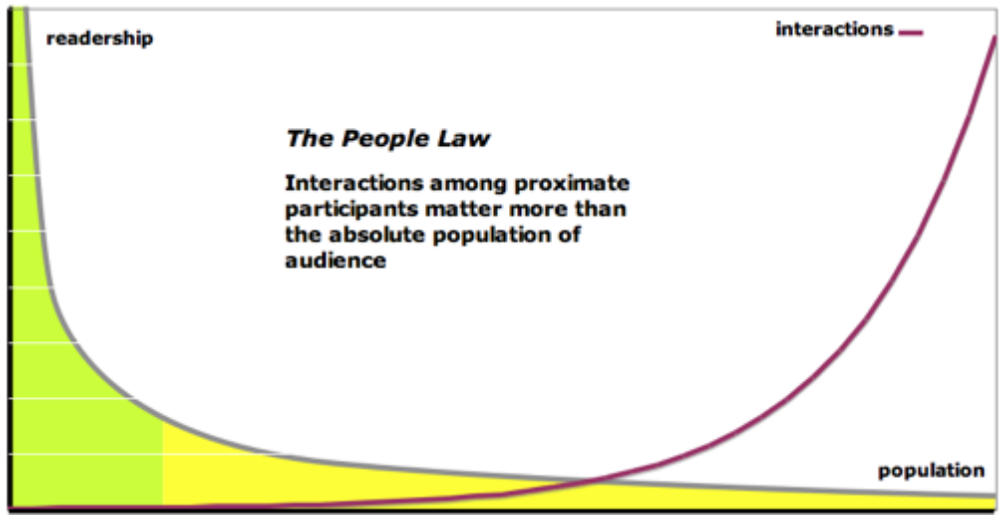

The problem with seeing the small groups in the long tail as the source of true collaborative knowledge creation was highlighted by Scoble and Israel's (2006) caution that online communities may become *echo chambers*, places where the high volume of conversation creates the illusion of *cognitive diversity* (i.e., the expression of ideas from diverse perspectives) but is really the repetition of a narrow range of similar ideas from the same group of people. This phenomenon has also come to be associated with *cultural tribalism*, a condition well described by Kitchin (1998):

> … communities based upon interests and not localities might well reduce diversity and narrow spheres of influence, as like will only be communicating with like. As such, rather than providing a better alternative to real-world communities cyberspace leads to dysfunctional on-line communities … (p. 90)

Cultural tribalism is thus portrayed as a highly probable equilibrium condition for all online communities. Since the cost of trial and switching are low, people will, over time, sample a large number of communities and gradually migrate to the ones wherein they feel most at home, those where they hear only what they want to hear. Such groups will be smaller, albeit close-knit. Conversation may be lively, but always among the same people, sharing a limited range of perspectives. Therefore, this migration to comfortable cognitive spaces should cause cognitive diversity, and the potential for collective knowledge creation, to be sacrificed.

Recent research (e.g., Page 2007; Surowiecki 2004) extols the virtues of cognitive diversity (i.e., a multiplicity of perspectives) in groups of all kinds: diverse thinking leads to better collaborative outcomes. Page (2007) recognizes a dual nature to diversity in cognition: diversity in preferences and diversity in thinking styles, or as he calls them, *cognitive toolboxes*. Page also notes there is some degree of interdependence between preferences and toolboxes since people develop the thinking skills needed to satisfy their preferences and then change their preferences based on their new ways of thinking (pp. 285-296). We are faced then with somewhat of a paradox: a diverse toolbox yields better outcomes, yet groups tend to self-organize based on preference similarity, thus limiting the thinking toolbox.

While preference similarity may be the primary basis on which to found a group, problem similarity is a factor that may help to alleviate the low cognitive diversity issue: people who share a problem that they cannot solve on their own are likely to seek each other out in an attempt to find a solution. People who come together in this manner may differ in the details of how they prefer their problem resolved, but those implementation details should not preclude a diverse group from forming to brainstorm possible solutions. Page (2007) also recognizes this situation:

> … diverse groups do perform better than homogeneous groups. And those situations in which they do perform better are far from random. … diverse groups perform better when the task is primarily problem solving, when their [differences] translate into relevant tools, when there is little or no preference diversity, and when their members get along with one another (p. 328).

Page's requirement for little preference diversity and interpersonal harmony can be relaxed if group members feel free and able to use the problem solution as a basis for satisfying their diverse preferences. In that situation, where no one anticipates a loss of satisfaction, then interpersonal relations should naturally be more harmonious. In the next section, past research into collective problem solving is related to collective learning and the argument is made that people with a common problem have an incentive to cooperate.

**Information Search and Learning**

Wilson (1999) proposed a problem-solving model of information seeking based on Kuhlthau's (1993) model of the stages of information seeking and Ellis' (1989) behavioral model of information searchers. Wilson's model sees problem-solving as the primary motivation of information seeking. His model envisions seekers acting out cycles of successive searches, each intended to reduce uncertainty but often increasing it. Even after the current problem is solved, people periodically re-enter the process to solve new problems as their needs or context change.

Problem-solving can be characterized as a learning activity. Even though learning is usually considered an individual activity, collective learning is sometimes preferable. Reynolds (1987) proposed *flocking theory* as a computational model that explains how the coordinated movement of a group can emerge from individuals making decisions based on personal information. Although flocking theory was developed as a solution to modeling the behavior of flocking birds and animals in computer graphics, Rosen (2002) proposed that flocking theory was a good explanation for self-organization in human social systems. He proposed that communication was the mechanism of cohesion in human society where a social network of individuals shares access to a collective body of knowledge that acts as a "roadmap" for coordinated action with little centralized control.

Axelrod (1984) described conditions under which normally self-seeking agents would cooperate when they perceived the benefits exceeded those of self-seeking. Johnson and Johnson (1988) developed that idea in the realm of education by describing the benefits of cooperative learning when individuals perceive their success to be entwined with group success. Their work may have been the first to describe a situation where maximizing group member heterogeneity was positively correlated with successful collective outcomes. By linking flocking theory with cooperative learning this study models a situation where individuals possess different aspects of the knowledge required to solve a mutual problem and thus have an incentive to cooperate.

In summary, idea markets help people satisfy preferences for ideas they like to entertain and also help them solve problems they hold in common. Commonality in idea preferences cause individuals to form long term idea communities, albeit with little diversity of perspective. These communities, however, will attract a regular influx of problem-motivated idea-seekers that increase the variety of perspectives, albeit briefly, as these solution-seekers probably only stay long enough to have their needs met. Although they may only stay briefly, the solution-seekers cause the knowledge possessed by the core group to expand by requiring them to think through problems that will undoubtedly vary to some degree. By satisfying both the needs of those with idea preferences and those with problems, idea communities emerge and become stable, increasing, long term stores of knowledge that attract a constant influx of diverse perspectives with the value of their collaborative encounters. Now we turn to a discussion of the means used in this study to model these idea markets.

**Agent-based Modeling as Theory Testing**

This study uses *agent-based modeling (ABM)* to simulate a population's search within an idea market. ABM is a simulation technique where a system is modeled as a collection of decision-making entities called *agents*. Agents typically move, interact and react based on a set of scripted behavioral rules. Some of the more advanced ABMs attempt to incorporate learning or evolutionary algorithms to make agent behavior increasingly better adapted to the circumstances of the simulation. However, even ABMs with simple agent behaviors can model complex interaction dynamics that are out of the reach of pure mathematical methods (Axelrod 1997).

ABM is also a way of thinking about modeling: it is a bottom-up methodology that describes a system as the outcome of individuals acting autonomously, rather than the result of system-wide laws that dictate individual behavior from the top down. As a result, it is easy to learn to use ABM because only the behavior of a single agent need be programmed and as many agents as desired can be replicated and set loose within the simulation. Many theorists (e.g., Bonabeau 2002; Casti 1997; Epstein & Axtell 1996) view ABM as providing the most natural way of modeling a system.

The results of an ABM simulation can often be unexpected and counterintuitive. These are attributes of *emergent* behavior, that is, behavior resulting from interactions between unsynchronized autonomous entities. Such interactions are often nonlinear, sensitive to initial conditions and stochastic. This aspect of ABM often makes the results difficult to explain. It is important therefore, to be sure agent behaviors and relevant environmental conditions are scripted in a manner grounded in strong theory and, if possible, consistent with empirical data. This three-legged stool is depicted in Figure 3, taken from Garcia, Rummel and Hauser's (2007) study of *history-friendly* agent-based models.

## FIGURE 3

## Framework for a History-Friendly Model

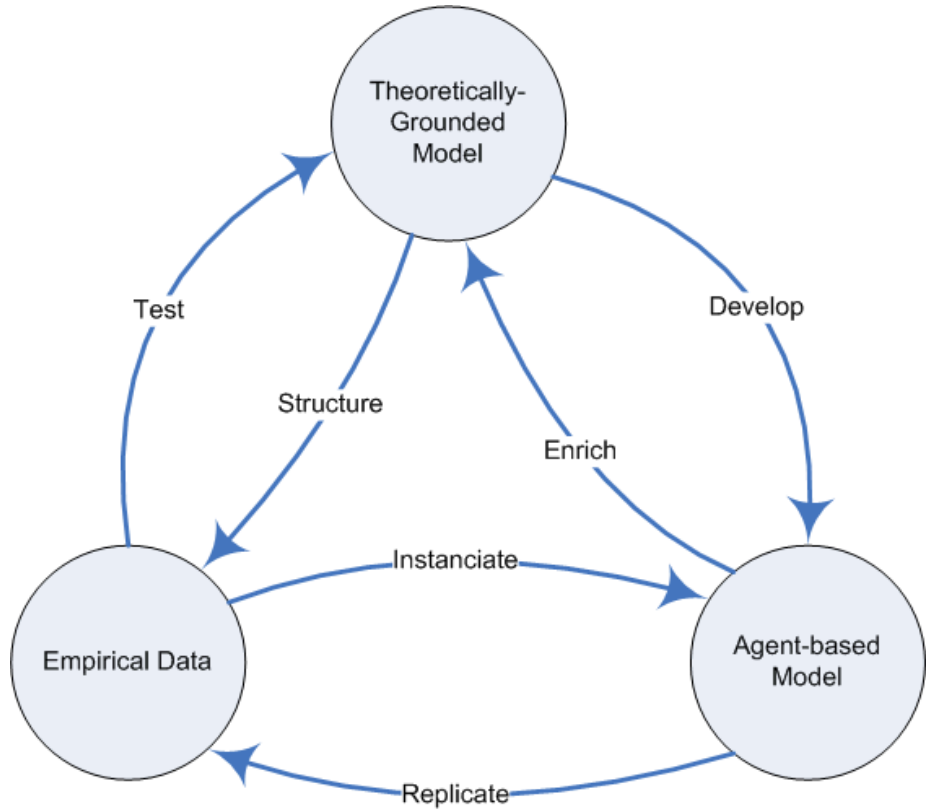

Carley (1996) observed that *validation*, the extent to which a model is true, can have more than one form, depending on the model's intended use. This study focuses on *theoretical validity*, the extent to which a model characterizes the real world and *external validity*, the extent to which simulated data matches real world data. Carley points out that most ABM research is obsessed with external validity as though it were the most important type of validity and its repeated demonstration critical to the future of ABM as a credible research tool. This emphasis on external validity does a disservice to ABM's powerful utility as a means of testing theoretical models by revealing what Krippendorf (1986) called a *behavior space*:

> The collection of behaviors a system can follow, the set of paths a system is capable of taking. A behavior space represents, sometimes graphically, and/or abstractly, and, often within many dimensions, just what a system can do so that what it actually does can be seen as a special case determined by initial conditions.

When assessing external validity the initial conditions and parameters of a model are tuned so as to reproduce a body of empirical data. Not only is any body of empirical data, like any single run of a simulation, a "special case determined by initial conditions," it is also a mixture of random noise and whatever effects are being measured. It is always hoped and assumed the effect is louder than the noise; however, we must guard against over-fitting a model to a dataset by adding more parameters than theory can justify. A better approach is to define the behavior space as a function of theory-based agent parameters, determine by simulation which parameters most influence the shape of the space and then locate the empirical dataset within the space. If the data does not fit the space, then the model is poorly specified. However, if the data fits within the space, we should rest content until we encounter a dataset that does not fit.

Purely analytical mathematical models are, like ABMs, based on theory-grounded parameters, yet their validity is more readily accepted. Edmonds and Hales (2005) argue that while ABMs should be seen as formal models, they lack the generality of pure analytical models. However, purely analytical models are invariably accompanied by simplifying assumptions that restrict their general applicability. Edmonds and Hales' point should be that the greater accessibility of ABMs makes them more susceptible for use (i.e., misuse) beyond their designed assumption limitations. It is wrong to label a methodology as inferior because it is easy to misuse. Rather, the focus should be on assessing how it was used in a research study and assess validity based on the consistency of its theoretical basis and the logic of its explanation for phenomena observed.

Thus far, the discussion has been abstract. As we move forward to discuss the model used in this study we will refer back to this discussion and be more specific.

## THE MODEL AND ITS IMPLEMENTATION

In this section, we look back to the research questions and describe an agent-based model and simulation scenario designed to build on the insights past research has granted. The scenario described below was implemented in Wilensky's (1999) *NetLogo* modeling environment (Version 4.0.4). A screenshot of the implementation user interface is given in Figure 4[2]. The simulation takes place within a 100 by 100 cell toroidal (i.e., doughnut-shaped) information space. Each coordinate within the space represents a theme, thematic similarity between points in the space varies inversely with the Euclidean distance between the points.

---

[2] The program file has been uploaded to OpenABM: http://www.openabm.org/site/model-archive/ideamarket

**FIGURE 4**

**Screenshot of Simulation User Interface Implemented in NetLogo**

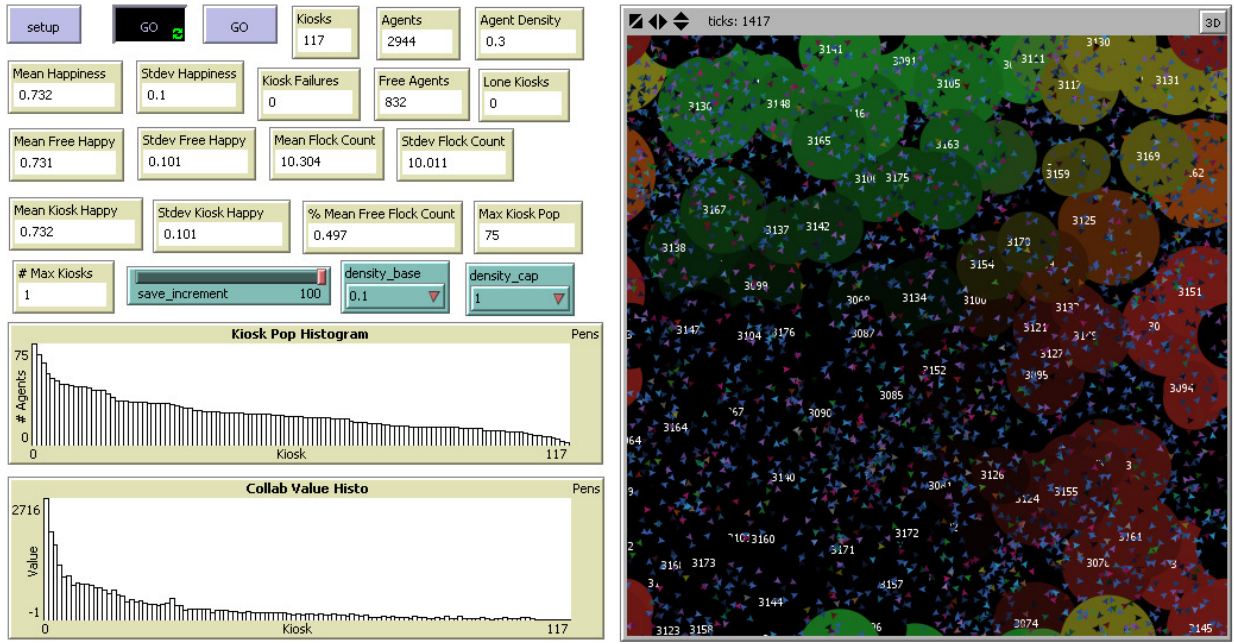

The space is populated with two classes of agent: information sources (or *kiosks*) and agents. Kiosks are stationary and are located at the coordinates of the theme for which they represent the intellectual home. Kiosks must be separated from each other by at least one cell within the space so they truly represent distinct themes. Now we return to the research questions:

a) **Is there a limit to the utility of cognitive diversity?** The simulation is reset to initial conditions and run ten times with agent population densities (agents per information space cell) ranging from 10% (1000 agents) to 100% (10,000 agents) in increments of 10%. Each run lasts 5000 *ticks*, NetLogo's time unit. These varied population levels and the long running interval are designed to allow some level of diversity satiation to emerge across a full spectrum of agent populations.

b) **Will an idea market facilitate the emergence of ideal levels of cognitive diversity?** A market fundamentally brings demand in balance with supply. At the beginning of a simulation run, an arbitrarily low, random number of kiosks (the supply component of the market) is created along with a relatively large number of agents (the demand component) randomly distributed throughout the information space at the density associated with the run. Using the mechanism described for the next question, this imbalance is relieved by an emergent dynamic that restructures the market, turning idea consumers into producers in a manner similar to entrepreneurship in economic markets.

c) **What causes ideal levels of cognitive diversity to emerge?** Agents begin wandering the space (Levy's 1937 algorithm) and soon start encountering other agents, comparing thematic preferences, flocking with similar others and looking for the kiosk closest to their thematic preference. As described in more detail below, agents unable to find preferred kiosks may become kiosks themselves and kiosks may revert to being agents if they cannot attract any agents with a matching preference. This dynamic of free migration between the states of thematic producer and consumer allows the optimal level of cognitive diversity to emerge that matches supply with demand.

d) **How can cognitive diversity be measured?** A mixed methodologies approach, where both quantitative and qualitative observations are made, was taken in this study to assess measures and patterns in the measurement of cognitive diversity. This study views cognitive diversity as a latent construct that, from a quantitative perspective, must be reflectively measured in a variety of ways (items a through d in the list below). It also proposes that valuable insights can also be gained by qualitatively comparing plots of measured values (items f through g in the list below).

   a. The steady-state number of kiosks across agent density populations.

   b. The steady-state proportion of kiosks to agents across agent density populations.

   c. The mean collaborative value received by agents across kiosks for each agent density.

   d. The mean number of agents per kiosk for each agent density.

   e. The standard deviation in Euclidean distances between a kiosk's coordinates and the thematic preference coordinates of its associated agents (this is a true direct measure of diversity unique to the simulation and unlikely to be portable to the real world).

   f. Qualitative comparisons of the shape of the distribution of agents across kiosks for each method of assessing a kiosk's attractiveness (discussed below).

   g. Qualitative comparisons of the shape of the distribution of collaborative value received by agents across kiosks for each method of assessing a kiosk's attractiveness (discussed below).

   h. Qualitative implications of the correlation between the number of agents associated with a kiosk and the total collaborative value received by these agents.

**Detailed Kiosk Attributes**

In addition to a theme, kiosks are endowed with the following attributes:

a) **Charisma.** Weber (1947) viewed charisma as a rare trait among individuals that is characterized by an uncanny ability to charm and persuade. This study models agents and kiosks as possessing of varying degrees of charisma. Highly charismatic kiosks can seduce passing agents away from their search and replace it with an interest in the theme espoused by the kiosk. Kiosk charisma is a random[3] number between 0 and 1. In a section below, various kiosk attraction scenarios are described. In those that exclude charisma, the seduction effect is not operant.

b) **Associated Agents.** As agents move through the space, kiosks come into their visual field. Agents assess the Euclidean distance between a visible kiosk's coordinates and their thematic preference coordinates normalized by the length of the information space diagonal (like Equation 2). If that normalized distance is less than 0.5, the agent will become associated with that kiosk until its attention span expires or a more "attractive" kiosk pulls it away. Other means by which an agent may be attracted to a kiosk are discussed below.

c) **Mass.** It is a well known phenomenon that crowds attract the curiosity of passersby. Berk (1974) explains this phenomenon in the first step of his *rational calculus* model of crowd action by noting that firstly, crowd members seek information. He points out that those recent arrivals to a crowd will begin by talking to other agents, probably asking: "What's going on here?" Thus, as partly explained in the discussion of charisma above and in greater detail below, this crowd attractiveness around a kiosk is one of six ways mass is modeled in this study. In that scenario, the charisma of a kiosk and the charisma of the agents associated with it, give a kiosk its *mass*, its attractive power, modeled as a gravitational fixed-point attractor, able to draw passing agents.

d) **Collaborative Value.** Gilder (1993) introduced the concept of *Metcalfe's Law*: "the systematic value of compatibly communicating devices grows as the square of their number." Bob Metcalfe's original statement of this "law," in a slide presentation sometime in the 1980's, referred specifically to communication technology like fax machines. Gilder expanded the definition by substituting "users" for "compatibly communicating devices," thus making Metcalfe's Law relevant to the context of social networks and collaboration. Odlyzko and Tilly (2005) argued that Metcalfe's Law overestimated the value of adding connections to a network by observing that not all connections in a network are equally valuable. They suggest the true value is better estimated by "n times the logarithm of n." This study, as a part of its mission, compares the results of modeling attractive mass (described above) as a function of crowd size with mass as a function of the collaborative value received by the agents associated with a kiosk and mass due to preference similarity. Odlyzko and Tilly's equation is used as the basis for calculating collaborative value (V) as shown in Equation 1.

(1) $$V = n_c \ln n_c$$

Where $n_C$ is the effective number of collaborators. Consistent with the theoretical discussion, the probability of two agents collaborating is inversely proportional to the distance between their thematic preference coordinates. However, the value of such collaboration varies directly with the distance between their thematic preference coordinates. Potential collaborators are filtered by likelihood of collaboration; then, the effective number of collaborators reflects the relative value of their collaboration as calculated by Equation 2.

---

[3] All random values are drawn from a uniform distribution unless otherwise stated.

$$n_C = \sum_{i \neq j} (d_{ij} / d_D) \quad (2)$$

Where $d_{ij}$ is the Euclidean distance between two collaborating agents' thematic preferences and $d_D$ is the length of the diagonal of the information space. Dividing the inter-agent thematic distance by the maximum distance in the space normalizes all inter-agent thematic distances between 0 and 1. The value of collaboration is thus higher when the agents are cognitively different, as represented by the distance between their thematic preferences[4].

**Detailed Agent Attributes**

Agents are mobile and represent seekers of information on a preferred theme. They are created with a random thematic preference expressed as a coordinate within the space, although the agent's behavior is not scripted to be aware the information space has coordinates. In addition to a preferred theme, agents are endowed with the following attributes:

a) **Charisma.** Like kiosks, agents are assigned charisma as a random number between 0 and 1. High charisma increases the probability an agent will become a kiosk if it cannot find a kiosk that matches its preferred theme. As described above, the charisma of agents crowding around a kiosk add to the attractive mass of a kiosk in one of the scenarios modeled.

b) **Attention span.** Attention span is modeled as a decrementing counter; initially set to a random value less than 1000. When the counter reaches zero any agent searching for a kiosk has a probability equal to its charisma of becoming a kiosk located at its preferred thematic preference coordinates. If there is already a kiosk at those coordinates, nearby locations, no closer than one cell from any existing kiosk are searched until an acceptable location is found.

c) **Impressionability.** When an agent is passing by a highly charismatic or high mass kiosk there is a probability equal to one minus its impressionability that it will be attracted to that kiosk and have its preferred theme replaced by that of the kiosk. Impressionability is a random value between 0 and 1. In simulation scenarios that do not use charisma, impressionability has no effect and agents change thematic preference only when their attention span expires.

d) **Range of vision.** Agents are given a circular field of vision with a radius randomly set between 5 and 10 cells around its current coordinates. Agents look for kiosks with similar theme and other agents with similar thematic preferences within their field of vision. Agents will crowd around kiosks or flock with any similar agents found.

---

[4] This may seem like "double accounting." Odlyzko and Tilly (2005) focus on network failures to interoperate, assessing collaborative value without considering diversity of perspective. Since that is one of this study's central tenants, diversity of perspective is piggy-backed on the factors they include. This approach may be open to the criticism of being overly conservative.

e) **Happiness.** Agent happiness is generally assessed as one minus the Euclidean distance between its physical coordinates and its thematic preference coordinates. In the next section, various models of kiosk attraction are discussed. One of those models, collaboration-value-based attraction, calculates happiness as the mean between the Euclidean distance just described and the collaboration value the agent receives normalized between 0 and 1. This change in the way happiness is calculated recognizes that idea market participants may forego happiness based on exercising their personal preferences so they can participate in helping their community, typically another source of happiness.

**Kiosks as Single-Point Attractors**

As agents move about the information space they are subject to the attractive forces of kiosks. This attraction force is modeled as a single point attractor exerting gravitational force proportional to its mass as described by Equation 3. An agent has a negligible mass of 1.

$$(3) \quad F = G \frac{m_1 m_2}{r^2}$$

Where F is the force of attraction, G is a constant (set here at 0.5), $m_1$ is the mass of the kiosk (already described, but discussed in greater detail below), $m_2$ is the mass of an agent (always 1) and r is the physical or thematic preference distance between agent and kiosk depending on the mass behavior scenario being run.

The simulations were run with six mass behavior scenarios:

1. **Crowd Attraction with Charisma.** As already mentioned, charisma is an indicator of individual attractiveness and crowds also have a magnetic quality, enhanced by the charisma of their members. The mass of a kiosk is therefore calculated as its own charisma added to the sum of the charisma of each agent associated with it.

2. **Crowd Attraction without Charisma.** The mass of the kiosk is simply the number of agents associated with the kiosk.

3. **Value Attraction with Charisma.** As already discussed, the total value agents receive from their collaboration with other agents associated with the same kiosk is a measure of the value of a kiosk's collaboration environment and of the kiosk itself. Here, the mass of a kiosk is calculated to be the sum of its charisma (as an initial, kick-starting attraction) and the total value received by the associated agents.

4. **Value Attraction without Charisma.** The mass of the kiosk is simply the total collaboration value received by the associated agents.

5. **Preference Attraction with Charisma.** The mass behavior scenarios described above all use aggregated measures (crowd size and total kiosk collaboration value) to set a kiosk mass that is uniformly applied to all passing agents. This scenario calculates a unique force of attraction based on the thematic distance between the kiosk and a passing agent: the closer the thematic distance the greater the attractive force. Mass is calculated as the sum of kiosk charisma and ten times one minus the thematic distance between kiosk and agent.

6. **Preference Attraction without Charisma**. Mass is calculated as ten times one minus the thematic distance between kiosk and agent.

## EMPIRICAL INVESTIGATION

In the discussion of agent-based modeling it was stated that one of the goals of this study was external validity, which is, comparing the simulation results with empirical data. While there are many idea markets, the blogosphere was selected for this study primarily due to the similarity of its structure to the simulation environment where content posted to a weblog (blog author entries and comments) are like kiosks and weblog readers like agents. Another selection factor was the ease of obtaining data concerning its growth characteristics, and the popularity and informational value ascribed to individual weblogs. On a quarterly basis, Technorati, a weblog search engine, publishes its *State of the Blogosphere* report containing a graph of the number of weblogs since March 2003. The graph from the latest report (Figure 5) is qualitatively compared with the pattern of growth in kiosks observed in this simulation (Sifry 2007).

## RESULTS AND DISCUSSION

**Is there a limit to the utility of cognitive diversity?**

Figure 6 indicates that the simulation showed consistent levels of satiation in the number of kiosks demanded by an agent population. Figure 7 (a) shows this satiation is robust across different population levels. Interestingly, Figure 7 (b) seems to show that the number of kiosks demanded per agent levels off to a constant proportion as the size of groups increase. This implies that the utility of cognitive diversity is higher in small groups where the probability of randomly assembling a maximized diversity of perspectives is lower.

**FIGURE 6**

**Kiosks Emergent over Time (Density = 0.1)**

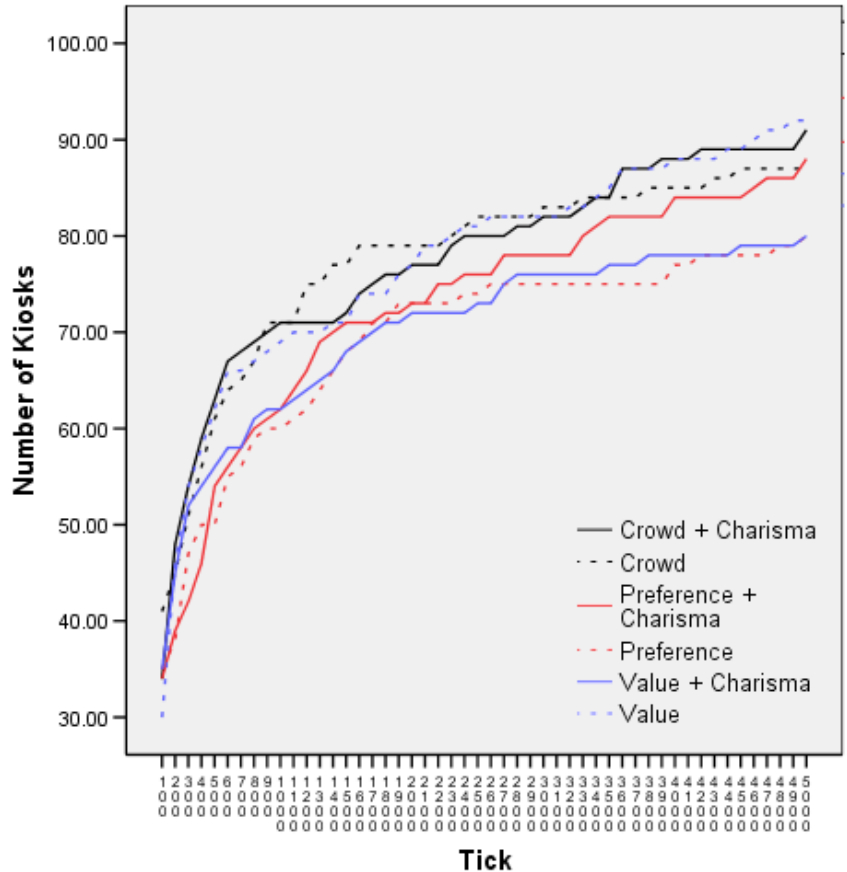

Figure 6 should be compared with Figure 5, the graph of the blogosphere's growth. It seems that the growth of the blogosphere has evolved to assume an s-shaped or logistic form, having passed an inflection point in its early exponential growth, its growth is slowing, trending toward a peak. The simulated model's behavior differs from the blogosphere in that the model's early kiosk growth rate was faster. The real world was slower to demand new weblogs than the model demanded new kiosks. However, once the inflection point was reached in the blogosphere's growth pattern, the patterns in the simulation and the blogosphere became similar.

It is interesting to note in Figure 6, that the satiation level differed across the ways in which kiosk attractive mass was modeled, most dramatically between collaboration-value-based mass and the other two ways of modeling attraction. If the simulation correctly models human behavior, the implication is that people forego the satisfaction of their preferences if the value they receive from collaboration is high enough and they perceive this value to be a primary goal. By foregoing individual preference the demand for diversity is reduced thus creating fewer kiosks. This may be modeling a situation where the wants of individuals are sacrificed for the good of a community.

**FIGURE 7 (a) and (b)**

**Kiosk Emergence Curves**

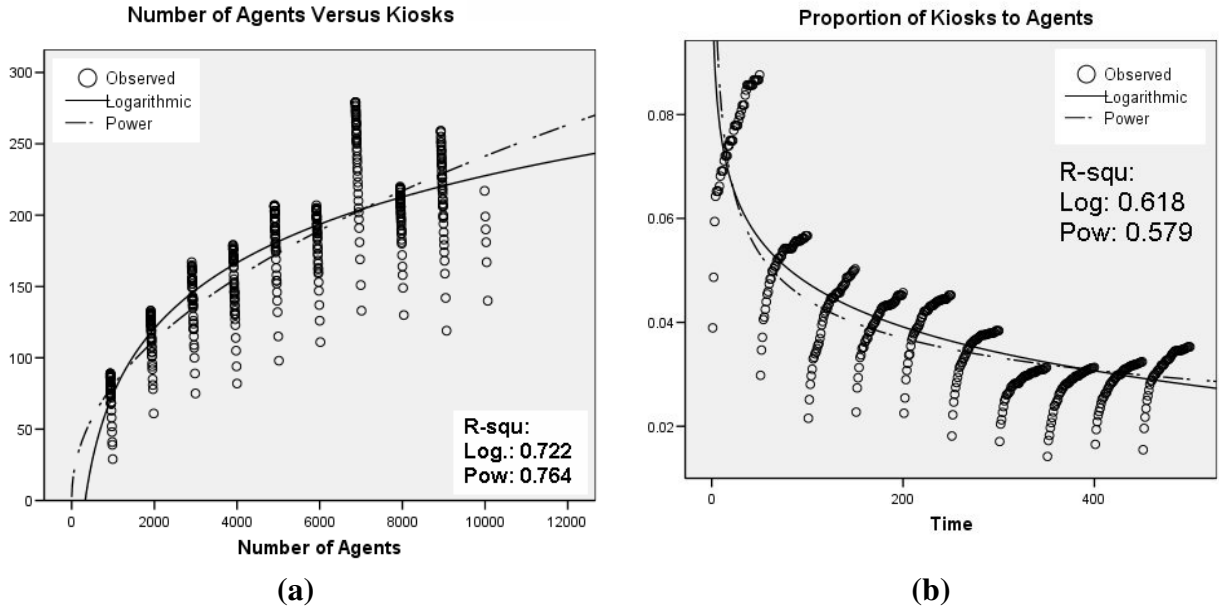

(a)  (b)

**Do idea markets cause ideal levels of cognitive diversity to emerge?**

    As already described, this study uses a direct measure of cognitive diversity tightly coupled to the design of the simulation: the standard deviation in Euclidean distance between a kiosk's coordinates and the thematic preference coordinates of its associated agents. Figure 8 shows a scatter plot of this cognitive diversity metric for kiosks ordered by population (descending) that is robust across densities and mass attraction models. Even though a poorly fitted regression line shows a slight decrease in cognitive diversity into the tail of the distribution, the mean level of diversity in each kiosk is very close to being uniform. The difference in cognitive diversity among the kiosks is really a matter of consistency: the more populated kiosks have greater consistency in diversity. A similar story is told by Figure 9, where the mean number of agents associated with a kiosk approaches the same value regardless of the mass attraction model used. It is also interesting to note from Figure 9 that a collaboration-value-based attraction model finds what seems to be an ideal kiosk population level faster.

# FIGURE 8

## Agent Cognitive Diversity across Kiosks

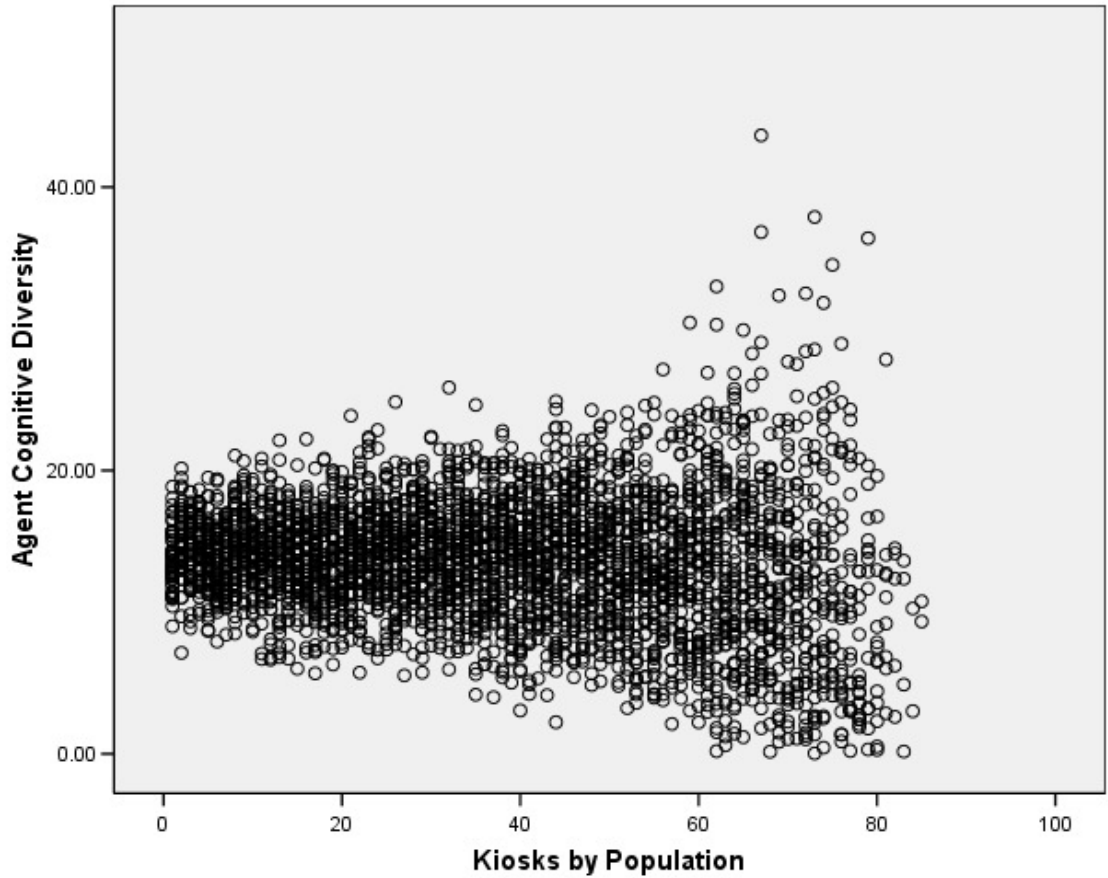

**Why does this simulation exhibit the power laws of real idea markets?**

      It was already mentioned that Shirky (2003) attributes the blogosphere's power law behavior to a scarcity of time among readers, an effect that may be facilitated (or exasperated) by search engines that tend to refer people to the most popular weblogs. The simulations in this study do not model time scarcity, they distribute preference uniformly among agents, and yet they display power-law-like behavior. This is probably due to the influence of modeling kiosks as single-point attractors, the influence of charisma and the use of flocking in the agent's search process, a conclusion consistent with Simon's (1955) preferential attraction mechanism. The interesting insight is that single-point attractors seem ubiquitous, not only in the realm of physics but in social systems as well.

**FIGURE 9**

**Mean Agents per Kiosk**

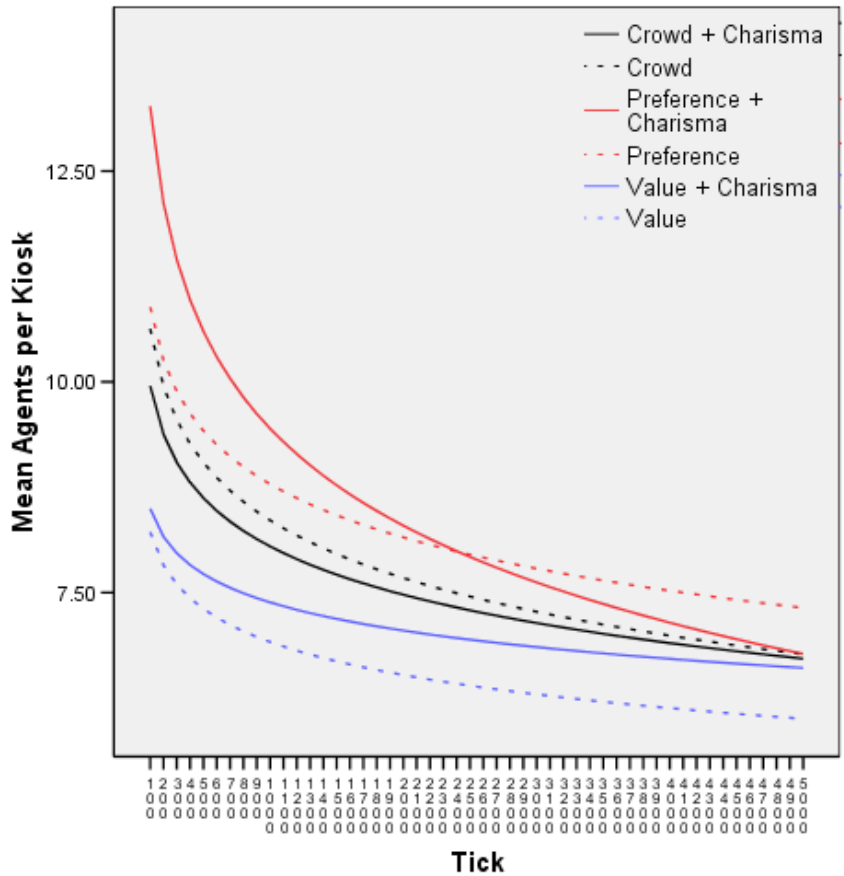

**Does preference-based attraction undermine cognitive diversity?**

      It was mentioned that Page (2007) recognized a dual nature to diversity in cognition: diversity in preferences and diversity in thinking styles and that groups tend to form based on preference commonality to the detriment of diverse thinking. This section presents evidence to suggest that the ubiquitous presence of single point attractors in idea markets cause diverse groups to form in spite of agents' thematic preferences. Figure 10 shows a power-law-like curve of the authority or value assigned to the top 100 most cited weblogs ordered in descending order of value. Figure 11 shows the same weblogs ordered by value, but indicates their popularity among Technorati members. The two value sets have a correlation of 0.686 ($\rho < 0.001$). While this indicates a moderate degree of similarity, the substantial difference justifies this study's differentiation between attraction due to value from that due to preference. Figure 10 echoes the pattern of Figure 1, the typical power law associated with the blogosphere. Figure 12 shows a similar pattern in collaboration value as modeled in this study, this general power law shape is characteristic of all collaboration value ordering of kiosks regardless of the mass attraction model used. So both the blogosphere and the simulation show preference diversity differs from thinking style diversity but not enough to be mutually exclusive.

**FIGURE 10**

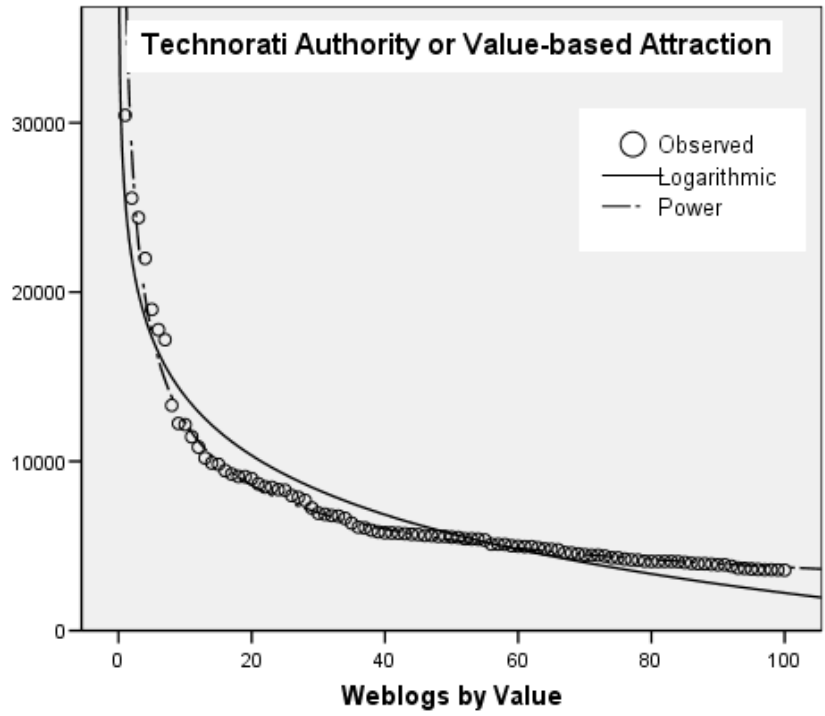

**FIGURE 11**

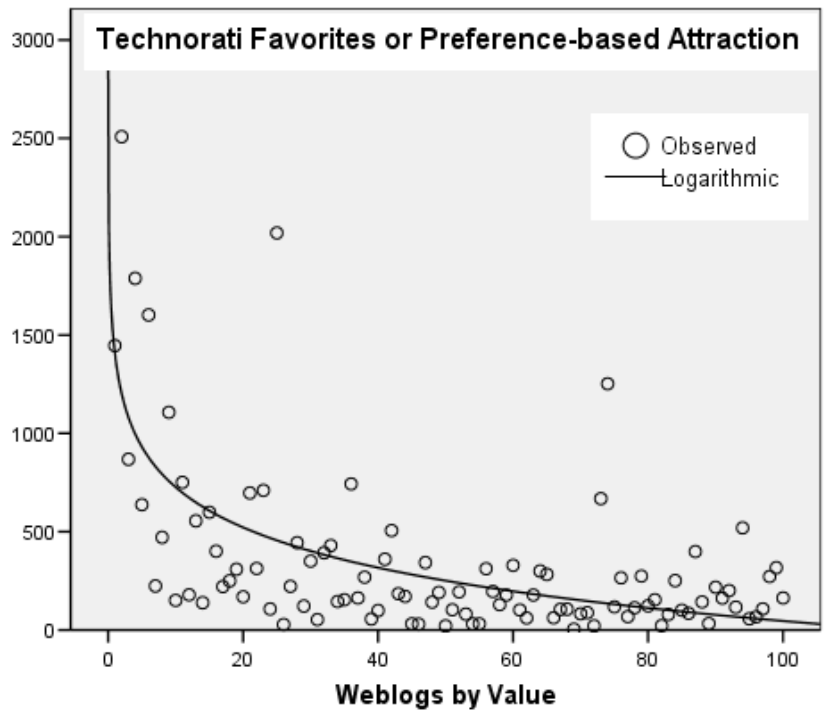

**FIGURE 12**

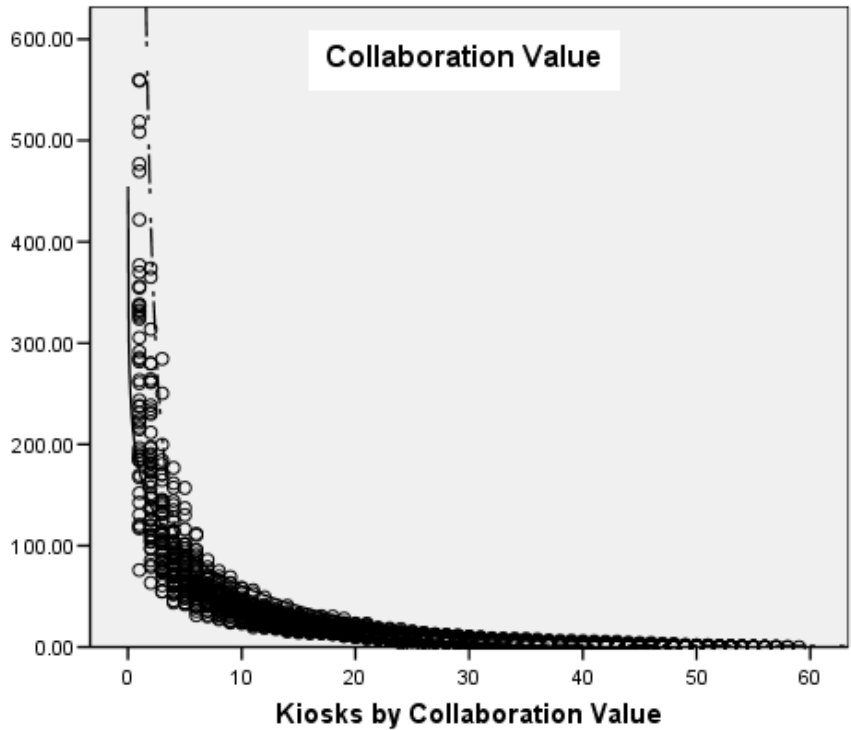

Figure 13 is a typical scatter plot of the happiness attribute values of agents associated with kiosks ordered by collaboration value. As already discussed, happiness is generally calculated from proximity to the preferred thematic location and agents were created with thematic preferences sampled from a uniform distribution. This explains why the model shows a near-uniform distribution of happiness. It is however, interesting to note that happiness levels are more consistent where collaboration values are highest than in the tail of the collaboration value curve. The dissimilarity between Figures 14 and 13 is loosely comparable to the difference observed in the real blogosphere between Figures 10 and 11. More striking is the similarity between Figures 11 and 13. The preference ratings of weblogs ordered by collaboration value were best-fit by a logarithmic curve (Figure 11) that approximated the shape of the power law in Figure 10, albeit with a lot of noise. In Figure 13, agent happiness was calculated as based on two factors: preference satisfaction and collaborative value received. Although Figure 13 has a more linear pattern, its negative slope is similar to the downward concave curve shown in Figure 11, implying that readers in the blogosphere may include the utility they receive from the authority of weblogs in their assessment of their favorite weblogs. So, although collaborative value and preference satisfaction are distinct in this study, these constructs may be related in the real world.

**FIGURE 13**

**Agent Happiness across Kiosks**

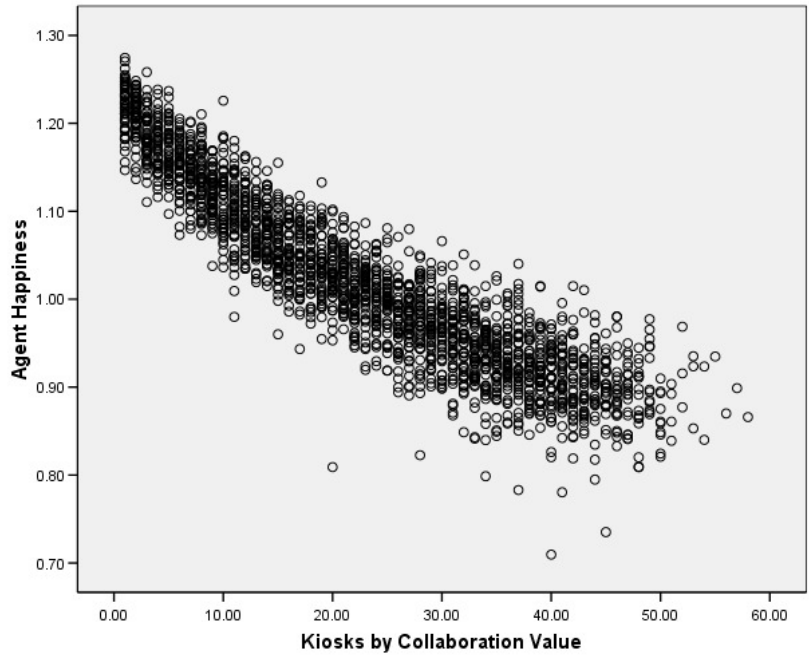

**FIGURE 14**

**Agent Cognitive Diversity across Kiosks**

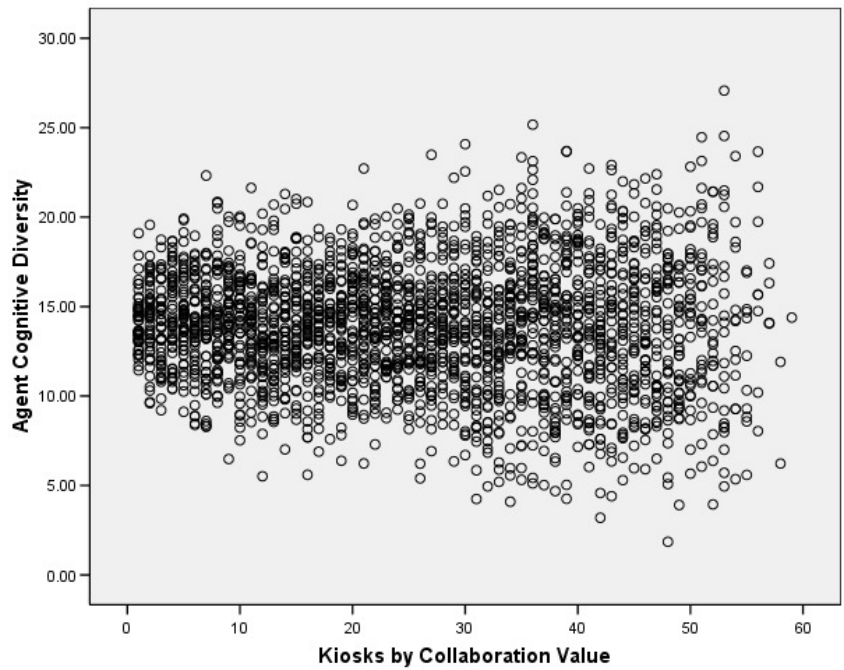

Let us revisit Blaser's (2007) speculations, depicted in Figure 2: interactions are richer in the tail of the blogosphere. We noted in the previous paragraph that the weblogs conferred with the highest collaboration value are often the most favored weblogs. In this study, the kiosks with the most agents are also the kiosks with the highest collaboration value (compare Figures 15 (a) and (b)); there is a 0.8 to 0.9 statistically significant correlation between these two value sets, robust across all agent densities and mass attraction models. Also, while Figure 8 shows a roughly uniform level of cognitive diversity across kiosks (also robust across densities and mass attraction models), the kiosks with the most agents have more consistent levels of cognitive diversity while the kiosks in the tail have some of the most diverse as well as some of the least diverse agent sets.

**FIGURE 15 (a) and (b)**

**Variations in Population and Collaborative Value across Kiosks**

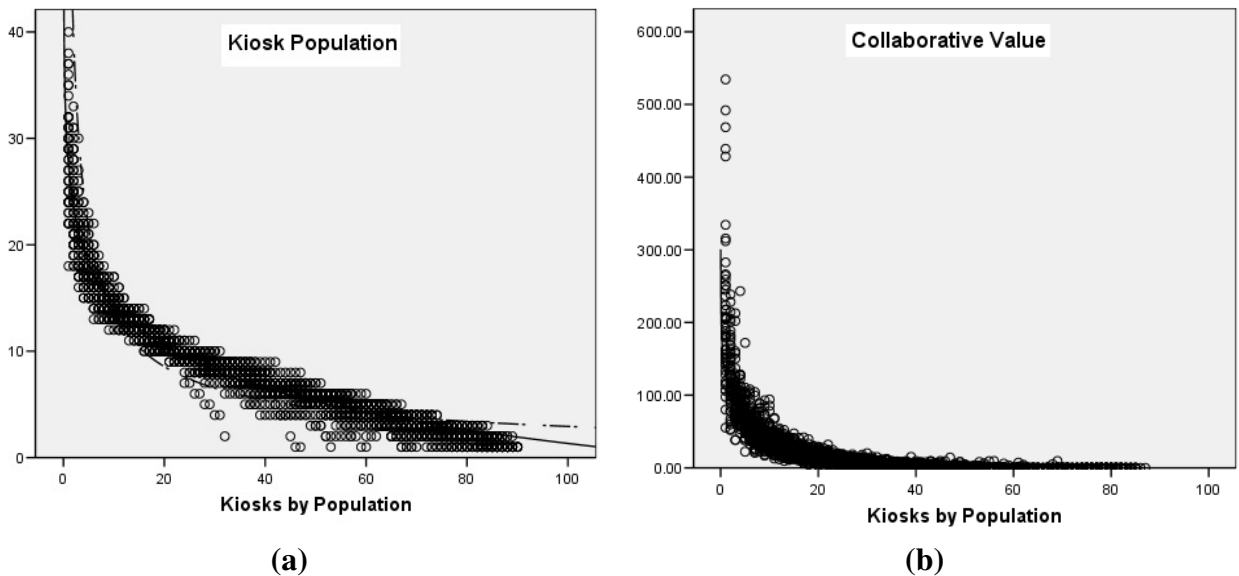

(a)  (b)

In Figure 14, the scatter plot of agent cognitive diversity across kiosks ordered by collaboration value revealed a similar pattern to Figure 8: consistent diversity among the kiosks with the highest collaboration value, less consistency among kiosks with lower collaboration value. If this simulation is accurately modeling the way real idea markets behave, and some evidence has been presented indicating it does, then Blaser's (2007) ideas about "The People Law" cannot be supported.

**Final Conclusions**

In this study, all the models of mass attraction yielded similar levels of cognitive diversity. However, as Figures 16 and 17 shows, collaborative-value-based attraction yielded lesser levels of collaborative value. This is attributed to the results shown in Figure 16, where collaborative-value-based attraction created smaller groups in the initial part of the simulation run. Both preference-based attraction and crowd attraction are easy to implement in the real world as the idea seeker always has knowledge of personal preference and can easily identify crowds that might call attention to something interesting. Collaboration-value-based attraction demands an extra interpersonal communication step where the seeker asks other seekers: How much does belonging to this group benefit you? The problem is further compounded by the need to assume all seekers value ideas in the same way in order to preserve some level of simplicity to the scenario. Crowd attraction may be a hard-wired adaptation that, in an organically simple way, brings people together around issues of general interest to ensure a high level of cognitive diversity will be present in order to make good decisions.

**FIGURE 16**

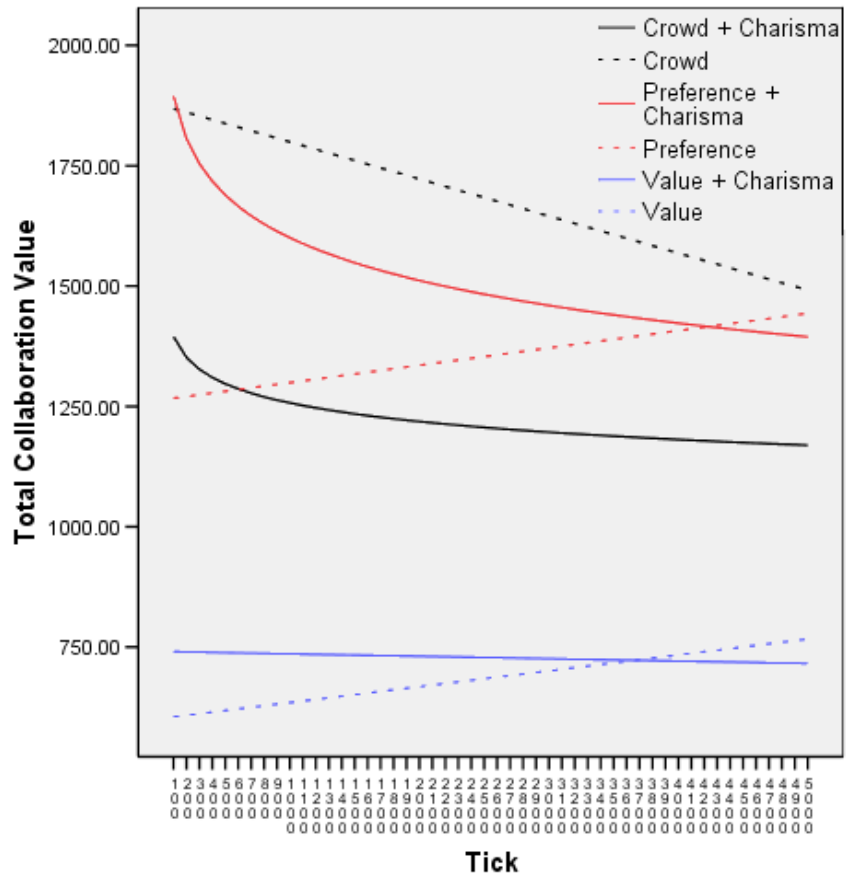

**Collaboration Value over Time**

**FIGURE 17**

**Collaboration Value across Kiosks (Value-Based Attraction)**

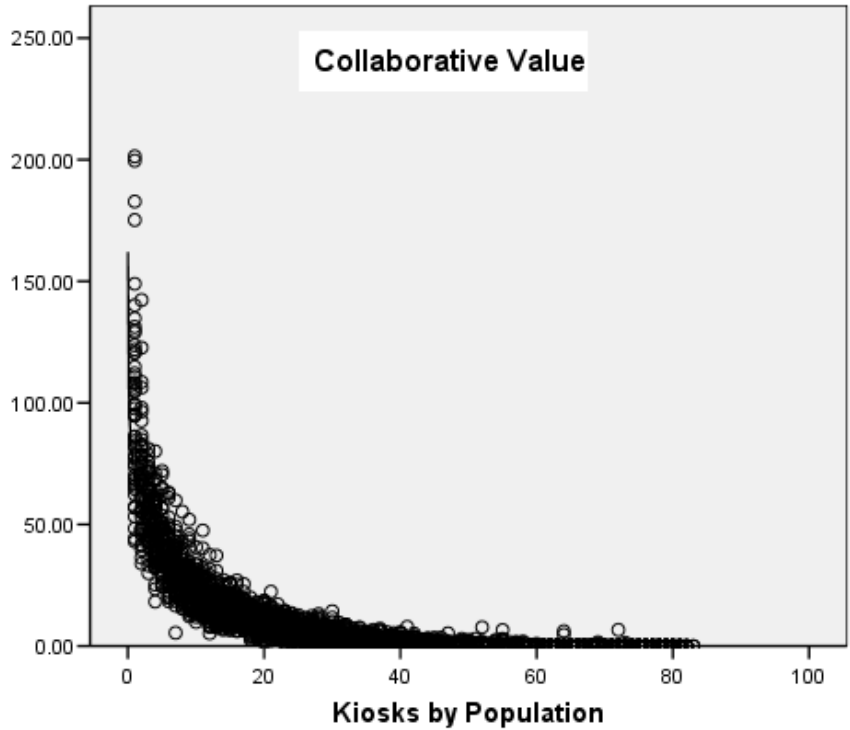

# DIRECTIONS FOR FUTURE RESEARCH

Previous research has found that the more people collaborate the more their mental models converge (e.g., Jeong and Chi 2007). Perhaps similar mental models homogenize problem-solving perspectives; so the longer people collaborate, the less the value of their collaboration. It seems that if thought diversity is better than uniformity there should be some sort of adaptive behavior that periodically or contingently perturbs situations where collaborative value has stagnated at minimal levels. Are individuals able to discern that the value of collaboration with some group has waned? Is that boredom? Does this realization reduce the attractiveness of a particular group, even to the extent of repulsion, driving people apart so they can, in a kind of reproductive isolation, develop their perspectives in diverging directions and thereby restore the potential for collaborative value? Or, is cognitive diversity maintained by continually shuffling group memberships, mingling individuals in a myriad of unique random combinations? How does comfort with the familiar, preference matching and boredom avoidance interact as behavioral forces?

# REFERENCES

Axelrod, Robert (1984), *The Evolution of Cooperation*, New York: Basic Books.

----, (1997), *The Complexity of Cooperation: Agent-Based Models of Competition and Collaboration* (Princeton Univ. Press, Princeton, NJ).

Berk, R. (1974), *Collective Behaviour*. Dubuque, IA: Brown.

Blaser, Britt (2007), *The People Law trumps the Power Law,* (accessed July 30, 2007), [available at http://www.blaserco.com/blogs/?cat=7]

Bonabeau, E. (2002), "Agent-based modeling: Methods and techniques for simulating human systems", *Proceedings of the National Academy of Science*, 99 (3), pp. 7280-7287.

Butler, Brian, Lee Sproull, Sara Kiesler and Robert Kraut (2002), "Community Effort in Online Groups: Who Does the Work and Why?" In S. Weisband and L. Atwater (Eds.) *Leadership at a Distance*, New York: Lawrence Erlbaum Associates.

Casti, J., (1997), *Would-Be Worlds: How Simulation Is Changing the World of Science* (Wiley, New York).

Carley, Kathleen M. (1996), "Validating computational models," Working Paper Carnegie Mellon University.

Edmonds, Bruce and David Hales (2005), "Computational Simulation as Theoretical Experiment," *Journal of Mathematical Sociology*, 29, 209-232.

Epstein, J. M. & Axtell, R. L., (1996), *Growing Artificial Societies: Social Science from the Bottom Up* (MIT Press, Cambridge, MA).

Ellis, D. (1989), "A behavioural approach to information retrieval design," *Journal of Documentation*, 46, pp. 318-338.

Gilder, George (1993), "Metcalfe's Law and legacy," *Forbes ASAP*, September 3.

Jeong, Heisawn and Michelene Chi (2007), "Knowledge convergence and collaborative learning," *Instructional Science*, 35, 287-315.

Johnson, Roger T. and David W. Johnson (1988), "Cooperative Learning: Two heads learn better than one," *Transforming Education*, 18 (Winter), p. 34

Kambil, Ajit (2003), "Betting on a New Market," *Trends and Ideas*, 1, 1.

Kitchin, R. (1998), *Cyberspace the world in the wires*, Chichester: Wiley.

Krippendorf, Klaus (1986), *A Dictionary of Cybernetics*, (accessed 26 Nov, 2009) [available at http://pespmc1.vub.ac.be/ASC/indexASC.html]

Kuhlthau, C. C. (1993). *Seeking meaning: a process approach to library and information services.* Norwood, NJ: Ablex.

Levy, Paul (1937), *Theorie de l'addition des variables aleatoires*, 2nd Edn, Gauthier-Villars, Paris (1954). [1st Edn (1937)]

Odlyzko, Andrew and Benjamin Tilly (2005), "A refutation of Metcalfe's Law and a better estimate for the value of networks and network interconnections", Working Paper University of Minnesota, (accessed Aug 9, 2007), [available at http://www.dtc.umn.edu/~odlyzko/doc/metcalfe.pdf]

Page, Scott (2007), *The Difference: How the Power of Diversity Creates Better Groups, Firms, Schools, and Societies*, Princeton, NJ: Princeton University Press

Reynolds, C. W. (1987), "Flocks, Herds, and Schools: A Distributed Behavioral Model", *Computer Graphics*, 21(4) (SIGGRAPH '87 Conference Proceedings), 25-34.

Rosen, D. (2002), "Flock theory: Cooperative evolution and self-organization of social systems", *Proceedings of the 2002 CASOS (Computational Analysis of Social and Organizational Systems) Conference*. Carnegie Mellon University, Pittsburgh, PA.

Scoble, Robert and Shel Israel (2006). *Naked Conversations*, Hoboken, NJ: John Wiley & Sons.

Shirky, Clay (2003), *Power Laws, Weblogs and Inequality,* (accessed May 8, 2007), [available at http://www.shirky.com/writings/powerlaw_weblog.html]

Sifry, David (2007), *The State of the Live Web*, (accessed Aug 11, 2007), [available at http://www.sifry.com/alerts/archives/000493.html]

Simon, Herbert (1955), "On a Class of Skew Distribution Functions," *Biometrika*, 42, 3/4, 425-440.

Surowiecki, James (2004), *The Wisdom of Crowds: Why the Many Are Smarter Than the Few and How Collective Wisdom Shapes Business, Economies, Societies and Nations*, New York: Doubleday.

Technorati (2007), *Popular Blogs*, (accessed Aug 11, 2007), [available at http://technorati.com/pop/blogs]

Warner, R. Stephen (1993), "Work in Progress toward a New Paradigm for the Sociological Study of Religion in the United States," *The American Journal of Sociology*, 98, 5, pp. 1044-1093.

Webb, Matt (2002), *Origin of Power Laws,* (accessed July 30, 2007), [available at http://interconnected.org/notes/2002/12/Origin_of_power_laws.txt]

Weber, Max (1947), *Max Weber: The Theory of Social and Economic Organization*. Translated by A. M. Henderson & Talcott Parsons. NY: The Free Press.

Wilensky, U. (1999), NetLogo. http://ccl.northwestern.edu/netlogo/. Center for Connected Learning and Computer-Based Modeling, Northwestern University. Evanston, IL.

Wilson, T. D. (1999), "Exploring models of information behaviour: the 'Uncertainty' Project," in: T.D. Wilson and D.K. Allen, eds. *Exploring the contexts of information behaviour: proceedings of the Second International Conference on Research in Information Needs, Seeking and Use in Different Contexts,* (pp. 55-66) London: Taylor Graham.